# Bending of light caused by gravitation: the same result via totally different philosophies

Tolga Yarman[1], Alexander Kholmetskii[2*], Metin Arik[3]

[1]Okan University, Istanbul, Turkey
[2*]Belarus State University, 4, Nezavisimosti Avenue, 220030 Minsk, Belarus, tel. +375 17 209 54 82, fax +375 17 209 54 45, e-mail: khol123@yahoo.com,
[3]Bogazici University, Istanbul, Turkey

**ABSTRACT.** We offer a concise and direct way to derive the bending angle of light (i.e. as generally called, "gravitational lensing"), while light grazes a star, through the approach suggested earlier by the first author, which is fundamentally based on the energy conservation law and the weak equivalence principle. We come out with the same result as that of the general theory of relativity (GTR), although the philosophies behind are totally different from each other. We emphasize that in our approach, there is no need to draw a distinction between light and ordinary matter, which makes our approach of gravity potentially compatible with quantum mechanics. Furthermore, our equation that furnishes gravitational lensing, also furnishes the result about the precession of the perihelion of a planet. The results obtained are discussed.

## 1. Introduction

By the beginning of 2013, our work about a new cosmological model, based on the law of energy conservation and the weak equivalence principle, was published; while in that paper, we postponed the formulation of our approach in a covariant form, we could yet predict the Hubble law, as well as an attractive answer to the dark energy quest (Yarman & Kholmetskii, 2013a); thus dark energy turned out to be the residue of a minimal positive acceleration of about $10^{-9}$ $g$ (where the symbol $g$ represents Earth's surface acceleration), of the initially exfoliating universe. Moreover, we had shown that at some earlier stages of the universe expansion, the acceleration could be negative.

The framework we then proposed was based on the first author's gravitation approach (Yarman, 2004; Yarman, 2006; Yarman, 2010a, b; Yarman, 2011; Yarman, 2013b), which we call hereinafter Yarman's approach, or in short YA, again essentially based on the law of energy conservation and accordingly, a re-formulation of the equivalence principle in a way, where the gravitational field energy may be a non-vanishing quantity in all possibly definable frames of reference. Just like it is the case yielded by the common general relativistic approach, in effect, we came out with the *"similitude"* or, say, *"likeness"* of gravitational and inertial masses, in the sense that the proper mass of any given object cancels out of the final equation of motion; this property is, as known, called "weak equivalence principle". At any rate, in our approach the proper mass $m$ of a massive object is altered by the gravitation field, so that its *overall motional relativistic energy E* is described by the relationship

$$E = \gamma mc^2 \left(1 - E_B/mc^2\right) \ , \tag{1}$$

where $\gamma$ is the Lorentz factor, associated with the motion of the object at hand; $E_B$ is the static binding energy, i.e. this is the energy required in order to bring $m$ quasistatically from where it sits, in the field, to infinity.

It should be noted that in a closed system $E$, owing to the law of energy conservation, must remain constant, the differential of which then furnishes the equation of motion.

Here we emphasize that the change of the rest mass of a particle embedded in the gravitational environment does not ultimately destroy the *similitude* about gravitational and inertial masses, we underlined above; therefore particles with different rest masses indeed happen to acquire the same acceleration in a given gravitation field.

Further we point out that the *similitude* of gravitational and inertial masses always allows choosing a reference frame, wherein the local geometry becomes pseudo-Euclidean. However, even in such a frame, the particle continues to *"sense"* the presence of the gravitational field due to the difference of its rest mass as referred to what it would have been in the absence of gravitation. In addition, as shown in refs. (Yarman, 2006; Yarman, 2010a, b; Yarman, 2011; Yarman, 2013), the change of the rest mass of the given particle induces a corresponding change of its temporal and spatial units, which thus occur to be a function of the static gravitational binding energy, i.e.

$$t = \frac{t_{empty\,space}}{1 - E_B/mc^2}, \quad r = \frac{r_{empty\,space}}{1 - E_B/mc^2} \quad . \tag{2}$$

Note that, because the static binding energy is proportional to *m*, these expressions do not depend on *m*.

These relationships signify that the resulting metric is *conformally flat* along with, in general, a non-constant conformal factor. We also observe that eqs. (2) do not contain any restrictions in extending the local geometry derived from these equations formally to the entire space-time. However, just like in any real situation, such a conformally flat metric field (extended to the entire space) constitutes an *abstraction,* and from an operational viewpoint, we can speak only about a local geometry in the given location for a given particle, unless information on the properties of other particles *(located in other spatial points)* is at hand.

Anyway, an infinitesimal spatial displacement of this particle in the gravitational field induces the variation of its static binding energy $E_B$ and the subsequent change of metric coefficients due to equations (2), applied to the co-moving frame of particle. Such a variation of metric in the co-moving frame versus *r* is "perceived" by the particle as the "gravitational force". Thus in the framework of YA, the overall change of metric of space-time in the frame co-moving with this particle represents rather a "secondary effect", caused by the change of the rest mass of particle in the gravitation field via the change of its static binding energy, but not as the direct influence of the gravitation field on the classically supposed space-time metric.

In this connection it becomes especially interesting to consider a number of effects of GRT (gravitational red shift, precession of perihelion of planet, bending of light, Shapiro delay), which are commonly considered as a direct manifestation of curved space-time in the presence of gravitational field (see, e.g. (Shapiro et al., 1968; Einstein, 1953; Feynman et al., 1977).

The gravitational red shift in the framework of YA has been considered in refs. (Yarman, 2004; Yarman, 2006) and also in Yarman's book (Yarman, 2010b). For a better feeling of the ideas presented herein, one may further refer to the following papers by the present authors (Yarman et al., 2007; Yarman et al., 2009; Yarman et al., 2011).

Both light bending and the precession of the perihelion of a planet can be derived from the same and one equation in YA, where one does not have to treat *light* and an *ordinary object* differently from each other.

It means, in particular, that a non-vanishing rest mass for the photon is postulated; anyway, as known, there have been several attempts to set an upper level to the rest mass of the photon (Alonso & Finn, 1968); lots of information about the photon rest mass can be already found in open sources (e.g. http://en.wikipedia.org/wiki/Photon#cite_note-amsler-28); its estimated numerical value in the framework of Yarman et al. is about $10^{-21}$ eV (Yarman et al., 2013c) and remains much less than the present limit of its measurement. (Note that we really do not need to know the particular value of the rest mass of the photon, in order to pursue the present approach.) Anyway, correspondingly, the propagation velocity *v* of a photon is just a bit less than *c* and, in general, depends on its frequency. However, the difference between *v* and *c* appears to be indistinguishable along with the modern experimental technique (Yarman et al., 2013c).

In the present paper we would like to concentrate on just the *light bending quest,* also relatedly Shapiro delay, leaving, digging more, in the precession of the perihelion of a planet to a future work.

We will show that light bending comes to be devilishly the same in both GRT and YA approaches, despite the fact that there exist insurmountable incompatibilities between the two theories. In fact, the same results are derived with respect to all main test problems both in GTR, and in YA, up to a third order Taylor expansion; in short, YA provided successfully all of basic predictions of GTR (Yarman, 2004; Yarman, 2006; Yarman, 2010a).

However, we stress here that there is no full conformity between the results of GRT and YA; in particular, YA yields a cosmological model quite different than that furnished by GRT, which suggests a solution of a number of long-standing problems of modern cosmology (Yarman & Kholmetskii, 2013a), and especially a solution to the dark energy problem. Of course, further development of YA approach is highly required (in particular, for its extension to the case of strong gravitation fields); in this aim, we are actually working on a covariant formulation of YA. At the same time, for a wider audience, we find very important to demonstrate in different and new ways that YA is a vital approach, and is fully compatible with all known test problems of GTR, including light bending.

For completeness, in section 2 we reproduce the derivation of light bending expression in the framework of GTR. In section 3, we reproduce, for readers' convenience, fundamentals and basic results of YA. Then in section 4, we provide an original derivation about light bending (gravitational lensing) in the framework of YA. We conclude in section 5. In effect, it is very interesting that light bending comes to be devilishly the same in both GRT and YA approaches, despite the fact that, as mentioned, there exist insurmountable incompatibilities between the two theories.

## 2. Light bending in GTR

Let us consider a ponderable mass *M* as the source of gravitational field, and choose the Schwarzschild metric, which can be written to the accuracy $c^{-2}$ in the isotropic form (see, e.g. (Lerner, 1997); here we use two-dimensional case, which is naturally applicable to the description of light bending):

$$ds^2 = (1 - 2\alpha)c^2 dt^2 - (1 + 2\alpha)(r^2 d\theta^2 + dr^2), \qquad (3)$$

where

$$\alpha = G_0 M / rc^2, \qquad (4)$$

$r, \theta$ are polar coordinates, $G_0$ is the gravitational constant, and *c* is the light velocity in vacuum.

We see that for the metric (3), lengths are contracted and periods of time are stretched, as referred to what they would have been, at a location free of gravitation in empty space.

Thus, a locally measured *(proper)* spatial interval $dl_L$, located at a distance *r* from the center of *M,* when seen by the distant observer, will be measured to be *dl,* so that

$$dl = dl_L / \sqrt{1 + 2\alpha}. \qquad (5)$$

In contrast, the periods of time are stretched. That is, the locally measured *(proper)* period of time $d\tau$, associated with a clock situated at *r,* when seen by the distant observer, is measured to be *dt,* so that

$$dt = d\tau / \sqrt{1 - 2\alpha}. \qquad (6)$$

Now consider an object in motion nearby $M$, at $r$, crossing the piece of trajectory $dl$ during the period of time $dt$, when measured by the distant observer. The coordinate velocity (observed by the distant observer) $v$ is, by definition,

$$v = dl/dt. \qquad (7)$$

The fixed local observer, situated at $r$, instead, will measure the velocity $v_L$, so that

$$v_L = dl_L/d\tau. \qquad (8)$$

The two velocities relate to each other via eqs. (5) and (6):

$$v = \frac{dl}{dt} = \frac{dl_L\sqrt{1-2\alpha}}{\sqrt{1+2\alpha}\,d\tau} \approx v_L(1-2\alpha) \qquad (9)$$

(to the adopted accuracy of calculations $c^{-2}$). This means that *proper velocities* slow down by the factor $(1-2\alpha)$, as seen by the distant observer.

*This holds for any velocity, thus also for the velocity* of light. Thereby, the proper velocity of light $c$, when measured in empty space, free of any gravitational field, slows down nearby a ponderable mass $M$, when assessed by a distant observer, and when passing by the altitude $r$, it becomes $c_L$, so that

$$c_L = c(1-2\alpha). \qquad (10)$$

The same result can be directly obtained from the fact that light travels along a null geosedic, where $ds^2=0$. Thence (in GTR) light is in effect, subjects to the *known refraction index*, leading to the result displayed by eq. (10). The occurrence delineated by eq. (10) is called, as well known, the *Shapiro delay* (Shapiro et al., 1968). Note that in GTR, the local fixed observer at $r$ would still measure $c$.

The foregoing reasoning, essentially based on the sameness of the effect of ordinary acceleration and the effect of gravitation (i.e. the principle of equivalence), led Einstein to the calculation of the deflection of light, thence amounting to twice that was originally predicted by the Newtonian approach. Along with eq. (10), Einstein (Einstein, 1953) originally used the Fermat principle (Feynman et al., 1977) to write for the deflection angle $\vartheta$ of light, grazing a star of the mass $M$ at the shortest distance $R$:

$$\vartheta = \int_{-\infty}^{\infty} \frac{|dc|}{c}\cotan\varphi = \int_{-\infty}^{\infty} 2|d\alpha|\frac{R}{z} = \frac{2G_0MR}{c^2}\int_{-\infty}^{\infty}\frac{dr}{r^2 z} = 4\frac{G_0M}{Rc^2}, \qquad (11)$$

where $\varphi$ is the angle drawn by the position vector $\boldsymbol{r}$ (extending from the centre of $M$ to the instantaneous location of the photon), and the direction of the impact parameter $\boldsymbol{R}$ (see Figure 1), and the quantity $r = \sqrt{R^2 + z^2}$ is the magnitude of the position vector $\boldsymbol{r}$.

Eq. (11) is well-known and can be found in books devoted to GTR. Though, not to be bothered with signs we preferred to use magnitudes of the given quantities.

## 3. Fundamentals of YA

Now, we will first shortly explain the fundamentals of YA. Before this, as conveyed, for readers' convenience, we would like to summarize some implications of eq. (1), which are presented in refs. (Yarman & Kholmetskii, 2013a; Yarman, 2004; Yarman, 2006; Yarman, 2010a, b; Yarman, 2011).

Suppose we have two celestial bodies interacting with each other, such as the Sun and a planet. We can thus assume (for simplicity but without any loss of generality) that one of them weighs infinitely more massive, as compared to the other.

Let $M$ be the mass of the Sun, and $m_{0\infty}$ the mass of the planet, but this, at infinity, where the gravitational effect vanishes. The restriction $M \gg m_{0\infty}$, we just framed for the masses, is not indeed a necessity for the approach we will sketch below (Sobczyk & Yarman, 2008); it is only a matter of convenience.

Our assumption thus makes that, when $m_{0\infty}$ is in motion around *M*, as regards to the distant observer, *M* always stays in place. Furthermore, the case we will handle herein, well fits in such a frame.

Now, consider that the planet of concern is engaged in its known motion around the Sun. This motion can be conceived as made of the *two gedanken subsequent steps:*

i) Bring the planet *quasistatically* from infinity to a given location *r* on its orbit around the Sun, but hold it there *at rest.*
ii) Deliver to it, at the mentioned location, its motion on the orbit in consideration.

Both of these steps can be assessed to be hypothetical, and they are. But afterall we are going to formulate an *energy quantity,* and we are based on our elementary knowledge that, the final result does not depend on how we sequence the anterior steps conceived to arrive at it.

The first step, as sketched by eq. (1) (which in our approach represents, in fact, the law of energy conservation, embodying the mass and energy equivalence), yields a decrease in the rest mass of $m_{0\infty}$, as much as the *static binding energy $E_B(r)$* coming into play (Yarman, 2004; Yarman, 2006; Yarman, 2010a). Thus $m_{0\infty}$ becomes $m(r)$, so that

$$m(r)c^2 = m_{0\infty}c^2 - E_B(r). \qquad (12)$$

Eq. (12) can be formulated as

**Law 1:** The rest mass, or the same rest energy, were the speed of light unity, of an object bound to a celestial body, in fact any given body it may interact with, amounts less than its rest mass measured in empty space, and this as much as its static binding energy vis-à-vis the gravitational source of concern.

Now, if one moves $m(r)$ quasistatically, from *r* to *r+dr,* thus as much as *dr,* he has to work against the gravitational attraction force, *M* exerts on $m(r)$; this then, owing to Law 1, will yield an increase in $m(r)$, as much as $dm(r)$, in such a way that

$$dm(r)c^2 = G_0 \frac{Mm(r)}{r^2} dr. \qquad (13)$$

The force term we make use of, is nothing else, but the usual *Newton gravitational attraction force.* However, we do not really borrow it from Newton (Yarman, 2006). To emphasize this novel consideration, we would like to state it as our next law.

**Law 2:** Were the spatial dependency $1/r^n$ of the Newton's gravitational attraction force $G_0 Mm(r)/r^n$, reigning between two masses $m(r)$ and *M*, *strictly at rest* with respect to each other, postulated to behave as such, along with an unknown exponent *n;* this exponent, must turn to be necessarily identical to the value of 2, in order to cope with the Lorentz transformations that would be observed by a distant observer, situated outside the gravitational field, in the case where the *static mass dipole* composed of $m(r)$ and *M* is brought to a uniform translational motion.

We have to emphasize that, if the masses $m(r)$ and *M* are not at rest with respect to each other, then, as Newton himself suspected, the attraction force between these masses is not anymore given straight by $G_0 Mm(r)/r^2$ (Yarman, 2004; Yarman, 2006).

The integration of eq. (13) yields immediately the rest mass of the bound object $m(r)$:

$$m(r) = m_{0\infty} e^{-\alpha}, \qquad (14)$$

where $\alpha$ is defined by eq. (4).[*]

---

[*] We like to point out that this result can be extended to the case of a *generic source mass,* such as a mass of any given shape, or a mass *M,* composed of the total mass of a given system, say, a galaxy, instead of having a point-like mass. Indeed, consider a total mass *M,* but whose shape is not spherical; the mass is not either distrbibuted

The comparison of Eqs. (12) and (14) furnishes the static binding energy $E_B(r)$ at the location $r$:

$$E_B(r) = m_{0\infty} c^2 \left(1 - e^{-\alpha}\right) . \qquad (15)$$

Along with the second step, we proposed right above, now we bring the planet (*already moved quasistatically from infinity to r*), to its orbital motion of concern, of velocity $v$, it is supposed to delineate at $r$.

This yields the multiplication of the rest mass $m(r)$ by the Lorentz factor $\gamma = \left(1 - v^2/c^2\right)^{-1/2}$, so that the overall relativistic mass $m_\gamma(r)$, or the same, the overall relativistic energy of the object, which should for an isolated system stays *constant* throughout in orbit, becomes (Sobczyk & Yarman, 2008)[†]

$$m_\gamma(r)c^2 = m_{0\infty}c^2 \frac{1 - E_B(r)/m_{0\infty}c^2}{\sqrt{1 - v^2/c^2}} = m_{0\infty}c^2 \frac{e^{-\alpha}}{\sqrt{1 - v^2/c^2}} = Constant \quad , \qquad (16)$$

---

uniformy, thus it bears the density $\rho(r')$ at the location $r'$, where $r'$ is the distance of the location of concern to the center of mass O, of $M$. Let us pick up an infinitesimal mass $dM(r')$, around $r'$ situated in the infinitely small volume $dV(r')$, around the location $r'$. Let us denominate by $m(r)$ the small rest mass, left out of the rest mass $m_{0\infty}$, when it is located at the distance $r$ from the center of mass of the generic mass $M$, if measured from the center of mass O, of this latter. We further designate by $V_T$ the total volume of the generic mass. We now move $m(r)$ as much as $dr$, and track this displacement by the vector $\underline{r}''$, whose origin is now the center of the infinitely small source mass $dM(r')$. Note that

$$dr = |d\underline{r}''| \cdot \qquad (i)$$

Let us further call $\beta$, the angle centered at O, and lying between $r'$ and $r$. We can then write

$$r''^2 = r'^2 + r^2 - 2rr' \cos\beta . \qquad (ii)$$

Under the circumstances, eq. (13) will be written as

$$\begin{aligned} dm(r)c^2 &= G_0 m(r) dr \int_{V_{Total}} \frac{\rho(r') dV(r')}{r''^2} \\ &= G_0 m(r) dr \int_{V_{Total}} \frac{\rho(r') 4\pi^2 r'^2 dr'}{r'^2 + r^2 - 2rr' \cos\beta} \end{aligned} . \qquad (iii)$$

In order to carry out the integration, it is appropriate to chose the direction of $r$, as the vertical axis of a spherical coordinate system, centered at O. The equator will be the plane perpendicular to r, at O. The angle $\beta$ can then be considered as the *polar-zenith angle,* where then $r'$ points to the infinitely small source mass $dM(r')$. We can further consider the azimuthal angle $\psi$, to define the projection of the location meant by $r'$ on the equator. Then the above equation will be written as

$$dm(r)c^2 = G_0 m(r) dr \int_{V_{Total}} \frac{\rho(r') \sin\beta \, d\beta \, d\psi \, r'^2 \, dr'}{r'^2 + r^2 - 2rr' \cos\beta} . \qquad (iv)$$

Anyway, Eq. (iv), via integration leads finally a result similar to what is given by eq. (14):

$$m(r) = m_{0\infty} \exp\left[-G_0 \int_r^\infty dr \left(\int_{V_{Total}} \frac{\rho(r') \sin\beta \, d\beta \, d\psi \, r'^2 \, dr'}{r'^2 + r^2 - 2rr' \cos\beta}\right)\right] . \qquad (v)$$

It will of course be interesting to handle the gravitational lensing based on such a general formulation, and compare the results, with the existing predictions, but the necessary load, certainly goes beyond the scope of this article.

[†] We recall that GTR predicts (see ref. 19, eq. (88.9))

$$m_\gamma(r)c^2 = mc^2 \frac{\sqrt{1 - 2\alpha}}{\sqrt{1 - v^2/c^2}} = Constant \quad , \qquad (i)$$

which coincides up to third second order of the corresponding Taylor expansion with Eq. (16). Note yet that the velocities in Eq.(i) are measured by a fixed local observer situated on the orbit. In any case, the striking similarity between the above equation and eq.(16) is to be underlined.

in the present approach; the *orbital velocity v* is the same, either for the distant observer, or the local observer on board of the planet. Such peculiarities are presented in ref. (Yarman, 2006).

The same essentially holds for the *speed of light,* which remains the same in YA, in a gravitational environment, when measured by a distant observer. This property is, however, not an assumption, but is well yeld by the quantum mechanical aspects of the present approach, and the reason is as follows. However, let us first recall that our distances and periods of time are altered just as much, via quantum mechanics ((Yarman, 2004; Yarman, 2006). It is that the *rest mass decrease,* via eq. (14) of the object in consideration in a field, it interacts with, when injected into its quantum mechanical description, well leads to size increase, and period of time stretching,[‡] by exactly the same amount, as referred to the distant observer in the macroscopic world. (This latter, in fact, is nothing else, but the *weakening of the internal energy* of the object in hand, thereby gravitational red shift).

Further, eq. (16), nailed to a constant, via differentiation leads to

$$-\frac{G_0 M}{r^2 c^2}\left(1-\frac{v^2}{c^2}\right)dr = d\alpha\left(1-\frac{v^2}{c^2}\right) = \frac{vdv}{c^2} \ . \qquad (17)$$

In YA, this equation, just like eq. (16), is valid for any object in a given trajectory, and does exclude any distinction between light and ordinary matter[§]. Yet, as elaborated above, light cruises - though, with a speed very close to the ultimate speed *c* - still, no matter how little, less than *c* (Yarman, et al., 2007).

---

[‡] And, this is exactly why the speed of light is, according to the present approach, not altered near a celestial body. Note however that, no matter what light is not affected in YA, it still takes a longer period of time to graze the ponderable body of concern, as referred to the distant observer, due to the stretching of lengths, which thence, well predicts the Shapiro delay, but through a totally different philosophy. Let us try to make this clear:

In GTR light slows down as much as $1-2\alpha$ [cf. eq. (10)], due to contraction of lengths by the amount of $\sqrt{1-2\alpha} \cong 1-\alpha$, and stretching of periods [cf. eqs. (5) and (6)] by the amount of about $1+\alpha$. The period of time, $\Delta t_{GTR}$ as assessed by a distant observer, light will take to graze a star, between the locations $z_i$ and $z_f$ sufficiently far away from the ponderable mass, along the direction *z,* is

$$\Delta t_{GTR} = \int_{z_i}^{z_f} \frac{dz\sqrt{1-2\alpha}}{c_0(1-2\alpha)} \cong \int_{z_i}^{z_f} \frac{dz}{c_0}\left(1+\alpha+\frac{3\alpha^2}{2}\right) \ ; \qquad (i)$$

note that *dz,* is the infinitely short distance along *z,* if there were no celestial body around, and we called $c_0$ the speed of light in empty space.

In YA, the speed of light is unaffected, yet, *quantum mechanically,* lengths are stretched as much as $e^{\alpha} \cong 1+\alpha$, and periods of time are stretched by practically the same amount as that predicted via the GTR (and this is why in YA the speed of light really remains unaltered). Thus in YA, the period of time, $\Delta t_{YA}$ as assessed by a distant observer, light will take to graze the given star, between the locations $z_i$ and $z_f$ still sufficiently away from the ponderable mass, along the direction *z,* is

$$\Delta t_{YA} = \int_{z_i}^{z_f} \frac{dze^{\alpha}}{c_0} \cong \int_{z_i}^{z_f} \frac{dz}{c_0}\left(1+\alpha+\frac{\alpha^2}{2}\right) \ . \qquad (ii)$$

So we land at practically the same result (though not identical). Nevertheless, here may be a possibility to check, which theory comes closer to the reality, if the accuracy provided by actual measurement techniques could ever allow it.

It is anyway evidently amazing that one arrives at practically the same Shapiro delay, yet through two totally different philosophies. Note further that one can easily generalize the formulation for a generic mass, the way we sketched in the footnote, below eq. (16) of the text.

[§] Precisely speaking, in YA, the gravitational consant is affected by the field (Yarman, 2011). This, via the application of the second equation of the set of eqs (2), brings in a factor of $e^{-2\alpha} \cong 1-2\alpha$, multiplying the classical Newton Force term, next to $(1-v^2/c^2)$. Since $\alpha$, for the light bending case is negligible as compared to $v^2/c^2$, herein we can confidently overlook the detail of concern.

# 4 Light bending in the present approach

In this section we will show how YA predicts the same gravitational lensing (light bending), as that predicted by the GTR, though based on a totally different philosophy than that of this latter theory. In YA, originally, also Fermat Principle was used, while geared though to derive the *extra deflection angle* drawn as compared to the Newtonian deflection (Yarman, 2006). Below, we will provide an even more powerful way, thus to obtain directly the overall light bending result through YA, which is based on the deflection undergone by the *deflection of the momentum vector of the photon, due to the gravitational pull,* and making clearer the mathematics and the physics behind it, thus allowing an easier grasp of the phenomenon in consideration.

The rigorous calculations of bending of light (or, as we will show elsewhere, of the precession of the perihelion of a planet, based on the *same and one* eq. (17)) in the framework of YA are tiring (just the way, in effect, the rigorous calculations of bending of light are, even in the classical Newtonian approach). Recall that originally Yarman had considered a perturbational approach to evaluate the bending of light (Yarman, 2006). Herein we will yet consider a *direct,* but rather simple, and concise calculation.

First of all, let us explain from the physical viewpoint, why the deflection angle for light, passing near the massive body *M*, should be larger in YA in comparison with that yielded by the classical Newtonian approach. It is that the given object in a gravitational field, for a *fixed velocity v*, assumes an itinerary with shorter instantaneous radii *(as compared to the corresponding Newtonian radii).* To begin with, recall that YA does not make any distinction between light and an ordinary object. Then consider eq. (17) for a case of a circular motion of an ordinary object. In this particular case this equation can be written as

$$\frac{G_0 M}{r^2}\left(1-\frac{v^2}{c^2}\right)=\frac{v^2}{r}, \qquad (18)$$

thus leading in YA approach to the orbit radius $r_Y$

$$r_Y = \frac{G_0 M}{v^2}\left(1-\frac{v^2}{c^2}\right), \qquad (19)$$

in comparison with the Newtonian orbit radius $r_N$:

$$r_N = \frac{G_0 M}{v^2}. \qquad (20)$$

The radius $r_Y$ yielded by YA (20) is clearly shorter than the Newtonian radius $r_N$ (19) for the given velocity *v*. Thence in YA, we can deduce at once that, one will have a more enhanced deflection for light grazing, say the Sun, along with shorter instantaneous itinerary radii, as compared to what he would come out with, based on Newton approach.

Note further that in YA the *overall relativistic mass* of the object is a constant, which we called $m_\gamma$ (see eq. (16)). The differential of the momentum vector $\boldsymbol{p} = m_\gamma \boldsymbol{v}$ of it, thus can be written in the simple form of $m_\gamma d\boldsymbol{v}$. This differential vector quantity, as known, is directed toward the center of attraction. Regarding the calculation of the bending angle, we need the component of $m_\gamma d\boldsymbol{v}$, perpendicular to the original direction of light. Given that the magnitude of $m_\gamma \boldsymbol{v}$ does not practically change throughout, the deflection angle for light, through an infinitely short itinerary, is then given by (see Figure 1)

$$d\theta = \frac{m_\gamma |d\boldsymbol{v}|\cos\varphi}{m_\gamma |\boldsymbol{v}|} = \frac{|d\boldsymbol{v}|\cos\varphi}{v}. \qquad (21)$$

Now let us demonstrate that eqs. (21) and (17), after we transform the latter into a vector form, are sufficient for the calculation of light deflection angle $\vartheta$ in the framework of YA to the accuracy $c^{-2}$, which is adopted hereinafter.

The simplest way to get the deflection angle appears to consider, as an artifact, two different beams of light, of magnitudes of velocities respectively $v_A$ and $v_B$, so that $v_B < v_A$, in any case, both being almost equal to $c$, but not exactly equal to it,[**] as they graze the Sun.

Due to the change in the momentum direction (see eq. (21)), both of the beams will be deflected practically as much as

$$d\vartheta = \frac{|d\mathbf{v}_A|\cos\varphi}{v_A} = \frac{|d\mathbf{v}_B|\cos\varphi}{v_B}. \tag{22}$$

Equal ratios can be written as the ratio of the differences of their respective numerators and denominators. Therefore, eq. (22) can as well be written as

$$d\vartheta = \frac{|d\mathbf{v}_A|\cos\varphi - |d\mathbf{v}_B|\cos\varphi}{v_A - v_B} = \frac{|d\mathbf{v}_A| - |d\mathbf{v}_B|}{v_A - v_B}\cos\varphi. \tag{23}$$

In order to determine the difference $|d\mathbf{v}_A| - |d\mathbf{v}_B|$, we further address to eq. (17). First we divide this equation side by side by the period of time $dt$, light takes to cross factually the distance delineated by the vector quantity $d\mathbf{r}$, and convert it into a vector equation (see, e.g. (Yarman, 2006; Yarman, 2010b):

$$-\frac{G_0 M}{r^2}\left(1 - \frac{v^2}{c^2}\right)\frac{\mathbf{r}}{r} = \frac{d\mathbf{v}}{dt}. \tag{24}$$

Then we equate the magnitudes of both sides of the above equation:

$$\frac{G_0 M}{r^2}\left(1 - \frac{v^2}{c^2}\right) = \frac{|d\mathbf{v}|}{dt}, \tag{25}$$

which yields:

$$|d\mathbf{v}| = \frac{G_0 M}{r^2}\left(1 - \frac{v^2}{c^2}\right)dt \approx \frac{G_0 M}{r^2}\left(1 - \frac{v^2}{c^2}\right)\frac{dz}{c} = \frac{2G_0 M}{r^2}(c-v)\frac{dr}{c}\sin\varphi = -2d\alpha(c-v)\sin\varphi. \tag{26}$$

We note that a direct usage of eq. (26) regarding the evaluation of $|d\mathbf{v}|$ in the calculation of the bending of the given light beam would be awkward due to the highly sensitive dependence of $r$ on $v$ (see eq. (19)). And this is exactly why we designed the artifact of calculation via the introduction of two different light beams, which allows us to get rid of any dependence of the deflection we aim to derive, on $r$ and $v$, simultaneously. Furthermore, here as conveyed, we have taken into account that $v$ is very close to $c$, so that

$$dt \approx \frac{dz}{c} = \frac{dr}{c}\sin\varphi, \text{ and } c^2 - v^2 = (c+v)(c-v) \approx 2c(c-v).$$

Let us now apply eq. (26) to both beams. Thus, $|d\mathbf{v}_A| = -2d\alpha(c - v_A)\sin\varphi$, $|d\mathbf{v}_B| = -2d\alpha(c - v_B)\sin\varphi$. Hence

$$|d\mathbf{v}_A| - |d\mathbf{v}_B| = 2d\alpha(v_A - v_B)\sin\varphi. \tag{27}$$

Substituting further eq. (27) into eq. (23), we obtain

$$d\theta = 2|d\alpha|\sin\varphi\cos\varphi. \tag{28}$$

Here again, we used magnitudes of the given quantities, to make clear that we have a deflection angle, defined positively.

---

[**] Recall that in YA, ligth travels in empty space with a velocity $v$ indistinguishably smaller than $c$, but it is still not exactly the ceiling $c$. The greater light frequency, the closer its speed is to $c$. YA, on the other hand, treats light, just like any other object, which makes that it is potentially compatible with quantum mechanics.

Therefore, the deflection angle of light is equal to

$$d\theta = \int \frac{2G_0 M}{r^2 c^2} dr \sin\varphi \cos\varphi = \int_{-\infty}^{\infty} \frac{2G_0 M}{r^2 c^2} dz \cos\varphi = \frac{2G_0 M R}{c^2} \int_{-\infty}^{\infty} \frac{dz}{\left(R^2 + z^2\right)^3} = 4\frac{G_0 M}{c^2 R}. \quad (29)$$

This terminates our novel derivation of the light deflection nearby a ponderable mass via our equation of motion, derived in the framework of YA.

It is interesting to note that, mathematically, the extra factor of 2, as compared to the result furnished by the Newtonian approach, comes from the decomposition of $c^2-v^2$ as $2(c-v)$, where $v$ is practically equal to $c$, and the factor $(1-v^2/c^2)$ coming to multiply the classical Newton force term in eq. (24), originates from the Lorentz coefficient taking place in eq. (16). So, the present derivation helps to grasp bot physically and mathematically the phenomenon in consideration.

## 4 Discussion

It is of course amazing that the result (29) we arrived at, finally, comes to be identical to the result obtained through the GTR (eq. (11)), though along with a totally different philosophy. Moreover, despite conservative reactions, we would still like to indicate a serious inconsistency in the classical derivation of eq. (11). Namely, observing that light is slowing down nearby a ponderable mass from $z=-\infty$ to $z=0$, and applying the Fermat principle, we obtain its deflection angle amounting to $\vartheta/2$. However, at the spatial domain corresponding to variation of $z$ from 0 to $+\infty$, the velocity of light is, according to the GTR, *increasing* with respect to $z$, and the application of the Fremat principle yields a deflection angle with the opposite sign. In other words, the integral (11) should be, rigorously speaking, separated into two parts taken with the reverse sign (i.e. $\left|\int_{-\infty}^{\infty}\right| \to \left|\int_{-\infty}^{0}\right| - \left|\int_{0}^{\infty}\right|$), and the overall deflection angle, if we choose to use the Fermat Principle, the way Einstein did in his book *(cf. Reference 9)*, must come out to be zero! Afterall light does not keep on slowing down, while grazing the star, all the way through. It slows down only between the itinerary $z=-\infty$ and $z=0$, and then, no matter what its speed remains smaller than that it bears in empty space, it keeps on accelerating between $z=0$ and $z=\infty$ (no matter how awkward this may look), until it reaches its speed in empty space. We can further refer to an optical analogy, where a light beam entering into some refractive medium (which we introduce to model the gravitational field nearby a ponderable mass, regardless how cumbersome may be the dependence of the refractive index throughout the medium), leaves the medium, as it comes out of it, at exactly the same spatial direction, as that of the original beam, and no overall deflection takes place; note please that when we say the *"same spatial direction"*, we mean the initial spatial direction of light to the medium, but now displaced, yet staying *parallel* to this, up to its exit location from the medium of concern. We have to recall at any rate that, despite the widely committed error, based on the misusage of the Fermat Principle in the calculation of the light bending, we just disclosed, here, we do not at all criticize the GTR, from this angle at all; after all the least action principle established within the framework of GTR and the ensuing geodesics, lead anyway to the measured value of the light bending, and in this article, as conveyed, we really do not propose to question the correctness of GTR. What we wanted to do yet, is simply to show that the usage of Fermat Principle *for the purpose of the derivation of light bending* within the framework of GTR, does not fit the nature of this principle.)

Note that in the framework of our approach, we did not address to the Fermat principle, by any means (no matter what the original algebraic set up furnishes anyway the expected result), but solely used the *momentum transfer to the light beam* due to the gravitational attraction toward the mass $M$ (see eq. (21)). It is obvious that in this approach the value of the deflection angle represents a monotonously increasing function in the entire space, whose integral yields eq. (29).

So we really stay tight within our *bare framework,* and are not in the need of bringing any other assumption, such as Fermat principle.[††]

Further on, we again stress that eq. (17) and its vectorial generalization (24) are equally applicable to an ordinary object and light. In particular, we will show in a subsequent paper that the same equation (24) also describes the precession of the perihelion of a planet.

It is even more important to stress that the identical treatment of an *ordinary object* and *light* in the framework of YA makes it fully compatible with quantum mechanics and opens a principal way toward the unification of both theories. As known, in quantum mechanics (at least in its non-relativistic version), due to wave-particle duality, the qualitative difference between massive particles and light photons practically disappears. Moreover, when we admit a finite (though tiny) rest mass of the photon (like in YA in some other theories (see, e.g. (Alonso & Finn, 1968; Yarman et al., 2013c), then in relativistic quantum theory the photons are described by the Klein-Gordon equation, which is equally applicable to any particle with even spin. In this case, the qualitative difference between light and particles is eliminated, too. This gives a key for the unification of YA gravity with quantum mechanics, which will be considered in a subsequent work.

---

[††] One may wonder, though, how come the algebra based on Fermat Principle setup embodies, still, led to the expected results. The answer is this: The alleged algebra, embodies the cotangent of the angle intercepting the actual location of the light photon at a distance r from the center of gravity, and the impact point on the star of radius R, in the integrand of concern, i.e. R/z. Whereas our algebra written along with the deflection due to the factual gravitational pool created by the ponderable mass, embodies the cosine of the same angle (cf. Reference 2), in the integrand coming into play, i.e. R/r. What is striking is that, the integration of both of the integrands in consideration, lead diabolically to the same mathematical result. Therefore, those who hoped to tap the correct bending results via (but incorrectly, as mentioned above), Fermat Principle, in fact, came unwillingly to tap a bending result due not to the slowing down of light photon nearby the ponderable mass, but a bending result, in fact, due to the pool exerted by this on the light photon. Anyway, the setups in question, no matter what they luckily reproduced the correct results, are after all, conceptually, incorrect. And we are pleased to get finally cleared this awkward situation.

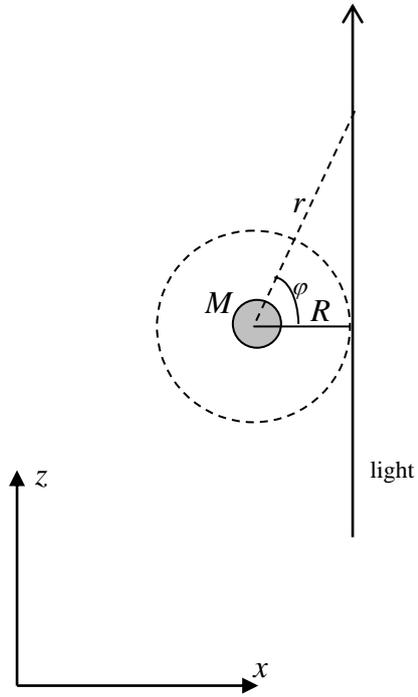

Figure 1. Propagation of light in a vicinity of ponderable mass *M*.